\documentclass [twocolumn, aps, showpacs] {revtex4}
\usepackage{graphicx}
\begin{document}
\draft
\title {Role of density imbalance in an interacting bilayer hole system}

\author {E. Tutuc, S. Melinte, E.P. De Poortere, R. Pillarisetty and M. Shayegan}
\address{Department of Electrical Engineering, Princeton
University, Princeton, NJ 08544}
\date{\today}
\begin{abstract}
We study interacting GaAs hole bilayers in the limit of zero
interlayer tunneling.  When the layers have equal density, we
observe a phase coherent bilayer quantum Hall state at total
filling factor $\nu=1$, flanked by a reentrant insulating phase at
nearby fillings which suggests the formation of a pinned, bilayer
Wigner crystal. As we transfer charge from one layer to another,
the phase coherent quantum Hall state becomes stronger evincing
its robustness against charge imbalance, but the insulating phase
disappears, suggesting that its stability requires the
commensurability of the two layers.
\end{abstract}
\pacs{73.50.-h, 71.70.Ej, 73.43.Qt}
\maketitle

When two layers of electrons are brought into close proximity so
that interlayer interaction is strong, new physical phenomena,
with no counterpart in the single layer case, can occur. Examples
include unique collective quantum Hall states (QHSs) at
even-denominator fractional fillings $\nu=1/2$ and $3/2$
\cite{suen-eisen}, and at $\nu=1$
\cite{murphy,lay,moon,spielman,kellogg2002} ($\nu$ is the total
filling factor of the bilayer system). These states are stabilized
by a combination of interlayer and intralayer Coulomb interaction.
The {\it bilayer} $\nu=1$ QHS is a particularly interesting state
as it possesses unique, spontaneous, interlayer phase coherence:
even in the limit of zero tunneling, the electrons spread between
both layers coherently. The state exhibits unusual properties such
as Josephson-like interlayer tunneling \cite{spielman} as well as
quantized Hall drag \cite{kellogg2002}.

Here we report magnetotransport experiments on interacting bilayer
{\it holes}, confined to two GaAs quantum wells with essentially
no interlayer tunneling, with an emphasis on the properties of the
system as the layer densities are made unequal (imbalanced). The
results of our study are highlighted in Fig. 1 where the
longitudinal resistivity ($\rho_{xx}$) vs. perpendicular magnetic
field ($B$) traces are shown. When the densities in the two layers
are equal (balanced), we observe a QHS at $\nu=1$, flanked by an
insulating phase (IP) reentrant around this QHS and extending to
filling factors as large as $\nu\approx1.1$ [Fig. 1(a)]. As we
transfer charge from one layer to another while maintaining the
total density constant, the IP at $\nu\approx1.1$ is destroyed
but, surprisingly, the $\nu=1$ QHS becomes stronger [Fig. 1(b)].
The weakening of the IP with charge transfer indicates that this
phase is stabilized by a delicate balance of interlayer and
intralayer Coulomb interaction, and suggests a pinned, {\it
bilayer} Wigner crystal. The strengthening of the $\nu=1$ QHS with
increasing charge transfer, on the other hand, demonstrates the
robustness of its phase coherence against charge imbalance.

Figure 1(c) reveals yet another intriguing feature of transport
coefficients in an imbalanced, interacting bilayer system:
$\rho_{xx}$ exhibits pronounced hysteresis at lower magnetic
fields near $\nu=2$, close to field values where either the
majority or the minority layer are expected to be at (layer)
filing factor 1. The hysteresis signals a first-order quantum
phase transition and is likely caused by an instability associated
with the layer index degree of freedom (pseudospin). We have
included the hysteretic data of Fig. 1(c) for completeness; we
will defer a discussion of this remarkable phenomenon to a future
publication. Here we focus on two main findings of our work,
namely, the robustness of the phase coherent $\nu=1$ QHS in
imbalanced bilyers and the IP that is reentrant around it.

We studied Si-modulation doped GaAs double-layer hole systems
grown on GaAs (311)A substrates. We measured six samples from four
different wafers, all displaying consistent results; here we
concentrate on data taken in three samples, from three different
wafers. In these samples, the holes are confined to two GaAs
quantum wells which have a width of 150$\AA$ each and are
separated by either a 110$\AA$ (samples A and C) or 75$\AA$
(sample B) wide AlAs barrier; sample C has typically a larger hole
density. We used patterned L-shape Hall bars aligned along the
[01$\bar{1}$] and [$\bar{2}$33] crystal directions. The hole
systems grown on GaAs (311)A substrates, including bilayer systems
used in this study, exhibit a mobility anisotropy stemming from an
anisotropic surface morphology, with the mobility being lower for
current parallel to [01$\bar{1}$]. Here we present data measured
with current along [01$\bar{1}$]; data taken with current parallel
to [$\bar{2}$33] are very similar. In all measurements the ohmic
contacts are connected to both layers. Metallic top and bottom
gates were added to control the densities in the layers. We
determined the layer densities from a careful examination of the
positions (in $B$) of the QHSs and the Hall coefficient, and from
the observation of a small step in the capacitance between the
bilayer and the top gate as the top layer is depleted. The
measurements were performed down to a temperature of $T=30$mK, and
using standard low-frequency lock-in techniques.

\begin{figure*}
\centering
\includegraphics[scale=0.7]{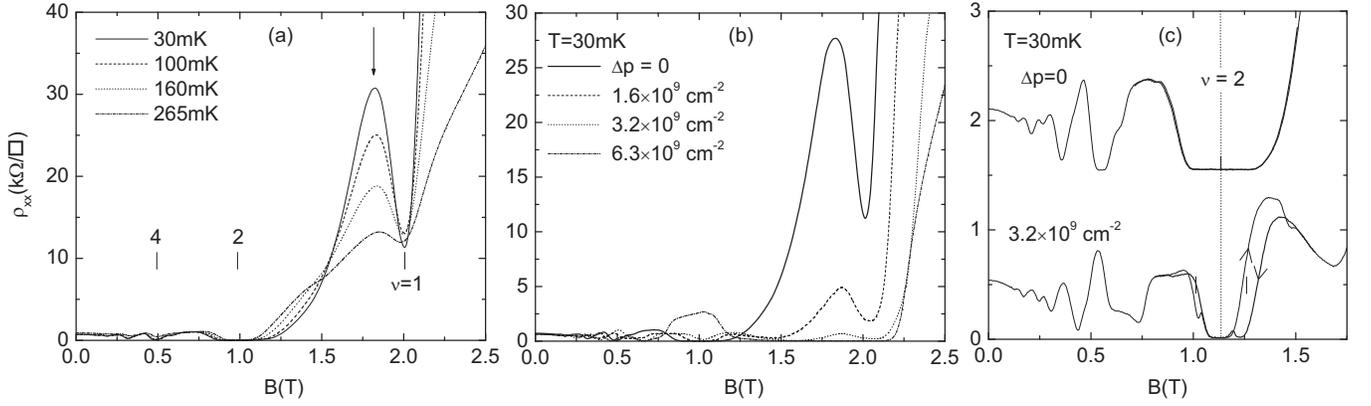}
\caption{\small{Resistivity vs. magnetic field traces for sample
A. (a) Temperature dependence of traces is shown; the total
bilayer density is $4.9\times10^{10}$ cm$^{-2}$, with both layers
having equal densities. The vertical arrow points to the
insulating phase developing near filling factor $\nu=1.1$. (b)
Data at $T = 30$mK for different values of the charge transfer,
$\Delta p$, while the total charge density is kept constant at
$4.9\times10^{10}$ cm$^{-2}$. (c) Data at $T=30$ mK at slightly
larger total density of $5.5\times10^{10}$ cm$^{-2}$, revealing
the development of a hysteresis in the magnetoresistance near
total filling $\nu=2$ when the bilayer is imbalanced. The upper
trace is shifted by 1.5 k$\Omega$ for clarity. The right (left)
tick mark in the lower trace indicates the expected position of
$\nu=1$ for the layer with larger (smaller) density.}}
\end{figure*}

In Fig. 1(a) we show $\rho_{xx}$ vs. $B$ traces for sample A,
taken at different temperatures; the total hole density is
$p_{tot}=4.9\times10^{10}$ cm$^{-2}$ and the two layers have equal
densities. Similar data are shown in Fig. 2 for sample B at
$p_{tot}=3.9\times10^{10}$ cm$^{-2}$. The $T=30mK$ data show a
strong minimum in $\rho_{xx}$ at total filling factor $\nu=1$, as
well as a developing plateau in the Hall resistance (Fig. 2).
Owing to the thick AlAs barriers, and the rather large hole
effective mass ($m^{\ast}=0.38m_{0}$, $m_{0}$ is the free electron
mass) interlayer tunneling is virtually zero \cite{tunneling},
implying that the observed $\nu=1$ QHS is stabilized solely by
interlayer coherence \cite{moon}. The ratio between the
interaction energy of carriers in different layers and in the same
layer is quantified by $d/l_{B}$, where $d$ is the interlayer
distance and $l_{B}=\sqrt{\hbar/eB}$ is the magnetic length at
$\nu=1$. For the cases examined in Figs. 1(a) and 2, this ratio is
1.45 and 1.12, respectively; these are close to $d/l_{B}$ for
which the coherent $\nu=1$ QHS was observed in earlier studies of
GaAs electron \cite{murphy,spielman} and hole \cite{hyndman}
bilayers.  Also consistent with previous results are our data (not
shown), taken on sample C, at densities of
$p_{tot}\geq9.15\times10^{10}$ cm$^{-2}$ ($d/l_{B}\geq1.98$),
which show no sign of a $\nu=1$ QHS.

\begin{figure}
\centering
\includegraphics[scale=0.35]{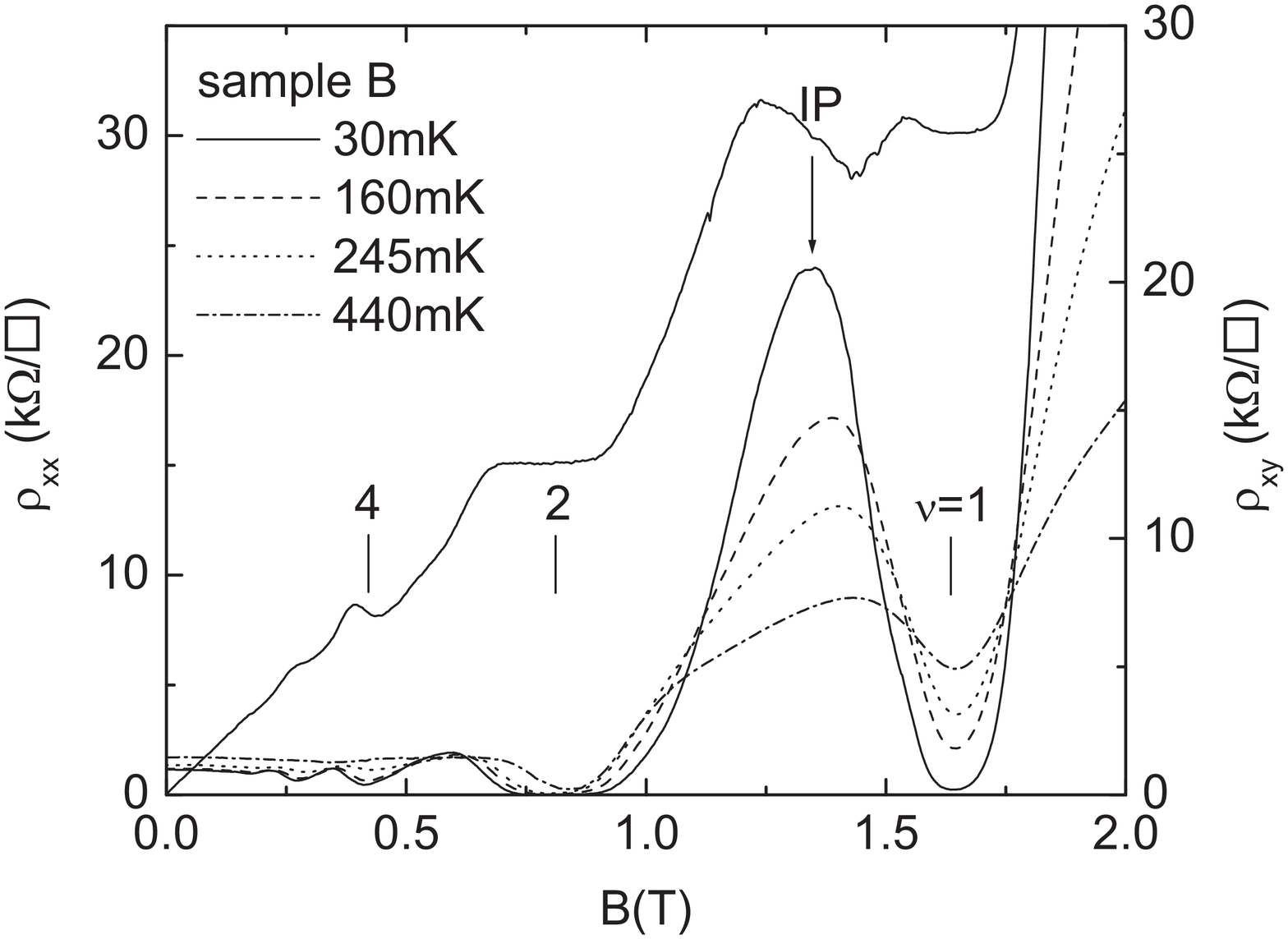}
\caption{\small{$\rho_{xx}$ vs. $B$ traces, at different
temperatures for sample B at a total density of $3.9\times10^{10}$
cm$^{-2}$, revealing a behavior qualitatively similar to Fig.
1(a). Also shown is the Hall resistivity $\rho_{xy}$ at $T=25$
mK.}}
\end{figure}

We now focus on the first major finding of our Letter, namely, the
behavior of the $\nu=1$ QHS as a function of charge imbalance
between the layers. In Fig. 1(b) we show $\rho_{xx}$ traces for
sample A, all taken at $p_{tot}=4.9\times10^{10}$ cm$^{-2}$
\cite{note2}, but with different densities in each layer. We
define the charge transfer from one layer to another as $\Delta
p=(p_{B}-p_{T})/2$, where $p_{B}$ and $p_{T}$ are respectively the
densities of bottom and top layers. The data of Fig. 1(b) reveal
qualitatively that the $\nu=1$ QHS is weakest when the layers have
equal densities and quickly becomes stronger as $\Delta p$
increases, i.e., when the bilayer system is imbalanced. We further
note that, while the results of Fig. 1(b) were taken by increasing
$p_{B}$ and reducing $p_{T}$, we obtain a similar set of traces by
transferring charge from the bottom layer to the top one.

\begin{figure}
\centering
\includegraphics[scale=0.35]{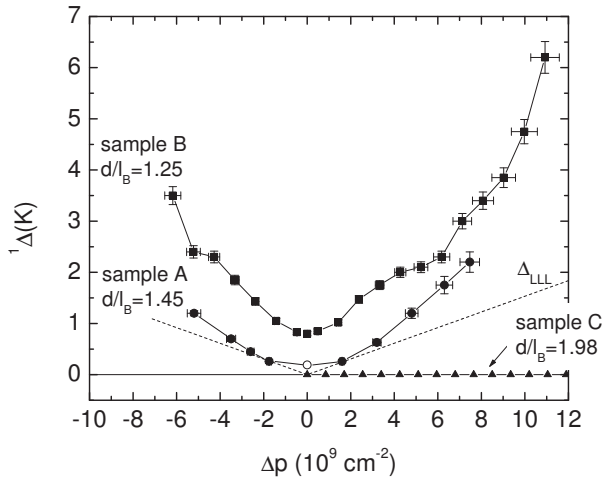}
\caption{\small{Energy gap ($^{1}\Delta$) of the $\nu=1$ QHS,
measured as a function of $\Delta p$, at
$p_{tot}=4.9\times10^{10}$ cm$^{-2}$ for samples A
($d/l_{B}=1.45$) and B ($d/l_{B}=1.25$). Sample C
($p_{tot}=9.15\times10^{10}$ cm$^{-2}$ and $d/l_{B}=1.98$) does
not exhibit a $\nu=1$ QH state in the range of imbalance studied;
this is indicated by a zero energy gap. The dotted lines show the
expected dependence of the separation between the lowest Landau
levels ($\Delta_{LLL}$) of the two hole layers on $\Delta p$ for
all three samples.}}
\end{figure}

To quantify the trend observed in Fig. 1(b), we have determined
the energy gap ($^{1}\Delta$) of the $\nu=1$ QHS from $\rho_{xx}$
activation measurements for both samples A and B at
$p_{tot}=4.9\times10^{10}$ cm$^{-2}$, corresponding to $d/l_{B}$
equal to 1.45 and 1.25, respectively.  We obtain the energy gap by
fitting an exponential dependence
$\rho_{xx}\propto\exp{(-^{1}\Delta/2k_{B}T)}$ to the $\nu=1$
resistivity vs. temperature data ($k_{B}$ is Boltzmann's
constant). We note that for sample A at $\Delta p=0$, a
determination of the energy gap at $\nu=1$ is impeded by the IP
developing at $\nu\approx1.1$. For this point only (shown in Fig.
3 as an open symbol) we have determined a pseudogap energy
\cite{pseudogap}. The measured $^{1}\Delta$ presented in Fig. 3,
quantitatively confirm the conclusion drawn qualitatively from the
data shown in Fig. 1(b): $^{1}\Delta$ is minimum at $\Delta p=0$
and increases when $|\Delta p|$ is increased \cite{note1}.

Our observation of a $\nu=1$ QHS in {\it balanced} bilayer hole
systems with a $d/l_{B}$ ratio of 1.45 or smaller is consistent
with previous reports for bilayer electron \cite{murphy,spielman}
and hole \cite{hyndman} systems. At such small layer separations,
the strong interlayer interaction leads to a $\nu=1$ QHS with
spontaneous phase coherence meaning that the charged carriers are
spread between both layers coherently, even in the limit of zero
interlayer tunneling. Our observation of a $\nu=1$ QHS in an
imbalanced bilayer carrier system in the limit of {\it zero
tunneling}, on the other hand, is new. As we discuss below, it
further confirms the phase coherent nature of the $\nu=1$ QHS and
demonstrates the robustness of this remarkable state against
charge imbalance.

Previous measurements of the $\nu=1$ QHS in imbalanced bilayers
have focused on electron systems with rather strong tunneling;
e.g., the tunneling energy $\Delta_{SAS}$ was $\approx6.8$ K in
Ref. \cite{sawada98}, and $\approx15$ K in Ref.
\cite{dolgopolov99}. For balanced states, the measured $\nu=1$ QHS
energy gaps measured in both of these studies were of the order of
$\Delta_{SAS}$. In Ref. \cite{sawada98}, an increase of
$^{1}\Delta$ with increasing charge imbalance was reported,
consistent with the expected increase in the subband separation
energy. In our hole bilayers, on the other hand, $^{1}\Delta$ is
determined by interaction and the resulting phase coherence, and
not by the subband separation. This is obvious for $\Delta p=0$,
where our estimated subband separation, $\Delta_{SAS}\leq1\mu$K
\cite{tunneling}, is orders of magnitude smaller than the measured
$^{1}\Delta$ (which is, e.g., about 1K for sample B). For finite
$\Delta p$, the subband separation in our samples, in which
interlayer tunneling is negligible, is essentially equal to the
energy separation between the lowest Landau levels of the two
layers, $\Delta_{LLL} = (2m^{\ast}/\pi\hbar^2)|\Delta p|$. Note
that this subband separation is independent of the layer
separation and therefore the same for all three samples. In Fig.
3, we have indicated $\Delta_{LLL}$ by dotted lines. An
examination of the data of Fig. 3 clearly shows that the energy
$\Delta_{LLL}$ is not what is leading to the stability of the
$\nu=1$ QHS is our samples. Indeed, at a given $\Delta p$, the
energy gap at $\nu=1$ varies from zero when $d/l_{B}$ is large
(sample C with $d/l_{B}\geq1.98$, in which no $\nu=1$ QHS is
observed in the range of imbalance studied here), to several K for
sample B, where $d/l_{B} = 1.25$ (Fig. 3). What determines the
strength of the $\nu=1$ QHS in therefore not the single-particle
$\Delta_{LLL}$ but rather the interaction and the ensuing phase
coherence.

In analogy with the real spin, the additional degree of freedom of
the electrons in bilayer systems is commonly described by a
pseudospin: if the electron is in the top (bottom) layer, then the
pseudospin is pointing up (down) along the $z$-axis, which is
defined by the growth direction. The stability of the phase
coherent $\nu=1$ QHS against charge imbalance can be understood in
a Hartree-Fock mean-field theory \cite{joglekar}, where it is
assumed that the pseudospin points in the same direction for all
electrons in the lowest Landau level. The favored direction will
be a result of the competition between the bias between the
layers, which tends to align the pseudospins along the $z$-axis,
and the capacitive energy cost, i.e., the Coulomb repulsion
between the layers, which forces the pseudospin to lie in the
$x$-$y$ plane (i.e. the plane of the 2D system). Moreover, thanks
to the exchange interaction, all pseudospins prefer to be aligned.
As a compromise, the pseudospin will lie in the $x$-$z$ plane with
the $z$-component increasing as the bilayer is further imbalanced.
It should be noted however, that according to reference
\cite{joglekar} it is also found that the gap of the $\nu=1$ QHS
is independent of the charge imbalance, in contrast to the results
of our measurements. Experimentally, we find that the gaps show an
approximately quadratic dependence on $\Delta p$ for small $\Delta
p$. In sample B, we also observe a "kink" in the dependence of the
gap on $\Delta p$ around $|\Delta p|=5\times10^{9}$ cm$^{-2}$; we
do not know the origin of this kink.

We now come to another important finding of our study, namely the
IPs observed near $\nu=1$. Data of Figs. 1(a) and 2 reveal an IP,
reentrant around the $\nu=1$ QHS, which extends to fillings as
large as $\nu\approx1.1$. Remarkably, a charge transfer of only a
few percent from one layer to another is sufficient to destroy
this IP. This conclusion is drawn from the $T$-dependence of
$\rho_{xx}$ at $\nu\approx1.1$, an example of which is shown in
Fig. 4 for sample B at $p_{tot}=4.9\times10^{10}$ cm$^{-2}$. It is
appealing to interpret the IP as a pinned, {\it bilayer} Wigner
crystal (WC) state. In previous studies of various low-disorder,
two-dimensional carrier systems, similar IPs, reentrant around
fractional QHSs at low Landau level fillings have been reported
\cite{shayegan}. Based on a variety of measurements and
calculations, such IPs are believed to be signatures of pinned WC
states, although so far there has been no definitive proof.

In the present case, it may appear surprising that a WC state
occurs at such high filling factors. However, this interpretation
is consistent with earlier experimental and theoretical studies
\cite{shayegan}, which suggest that, in interacting bilayer
systems a WC state with commensurate layer lattices can be
stabilized at filling factors higher than in the single-layer
case. For example, while a single-layer GaAs electron system
displays a reentrant IP (interpreted as a pinned WC) near
$\nu=1/5$, an interacting GaAs bilayer electron system exhibits a
similar IP around $\nu=1/2$ \cite{hari}. Experimental studies have
also revealed that single-layer, dilute, GaAs hole systems exhibit
an IP near $\nu=1/3$, a higher filling factor than the
corresponding single-layer GaAs electrons ($\nu=1/5$); this is
also consistent with the WC picture: thanks to the Landau level
mixing facilitated by the larger hole effective mass, the WC state
is favored at higher fillings \cite{shayegan}.  Based on these
observations, the formation of a WC state reentrant around $\nu=1$
in a bilayer 2D hole system with strong interlayer interaction
appears plausible.

\begin{figure}
\centering
\includegraphics[scale=0.35]{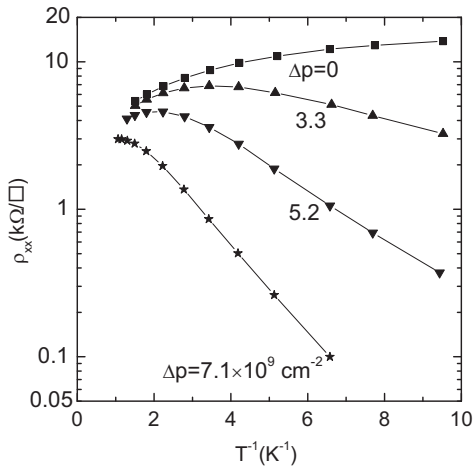}
\caption{\small{Temperature dependence of $\rho_{xx}$ at
$\nu\approx1.1$ (the IP), measured at several values of $\Delta p$
in sample B. The total density is $4.9\times10^{10}$ cm$^{-2}$.}}
\end{figure}

The data of Fig. 4 reveal an intriguing feature. When the bilayer
system is balanced, $\rho_{xx}$ monotonically increases with
decreasing $T$. For sufficiently large imbalance ($\Delta
p=7.1\times10^{9}$ cm$^{-2}$), the decrease of $\rho_{xx}$ with
decreasing $T$ is also monotonic. When only slightly imbalanced,
however, $\rho_{xx}$ at $\nu\approx1.1$ initially increases as $T$
is lowered but then at the lowest temperatures it decreases
\cite{tmax}. This non-monotonic $T$-dependence is unusual and may
be caused by a competition between the WC state and the nearby
phase-coherent $\nu=1$ QHS. Indeed for $\Delta p=7.1\times10^{9}$
cm$^{-2}$, the $\nu=1$ QHS is quite strong and, at the lowest
temperatures, exhibits a $\rho_{xx}$ minimum that extends to
$\nu\approx1.1$. This may account for the very strong
$T$-dependence of the $\nu\approx1.1$ $\rho_{xx}$ minimum at
$\Delta p=7.1\times10^{9}$ cm$^{-2}$ \cite{strongTdependence}.

Our observations highlight the rich and subtle nature of many-body
states that exist at nearby fillings in GaAs bilayer hole systems
in the limit of zero tunneling but with small layer separation.
When balanced, such systems show a phase-coherent $\nu=1$ QHS
state, flanked by a reentrant IP, possibly a pinned bilayer WC
state that is stabilized at such high fillings because of
interlayer interaction. When the bilayer is imbalanced, the
$\nu=1$ QHS state survives and even gets stronger, consistent with
its phase-coherent origin. The IP, on the other hand, disappears,
suggesting that its stability requires commensurability (of the WC
lattices) in the two layers.

We thank A.H. MacDonald, F.D.M. Haldane and Y.P. Shkolnikov for
fruitful discussions, and NSF and DOE for support.


\begin{thebibliography}{10}
\bibitem{note2}
The total density is kept constant within 3\%. The charge transfer
$\Delta p$ is measured with an accuracy of 6\%.
\bibitem{note1}
The slight asymmetry of $^{1}\Delta$ with respect to $\Delta p$ in
Fig. 3 is likely caused by unintentional differences between the
top and the bottom quantum wells.
\bibitem{tmax}
A plot of  $T_{max}$ vs $\Delta p$, where $T_{max}$ is defined as
the temperature at which $\rho_{xx}$ vs. $T$ at $\nu\approx1.1$
reaches a maximum, shows that $T_{max}$ depends linearly on
$\Delta p$ and extrapolates to zero as $\Delta p$ approaches zero.
\bibitem{strongTdependence}
The dependence in fact appears to be activated, although the
activation energy is only about 1/3 of the activation energy
measured for the nearby $\nu=1$ QHS at this $\Delta p$.
\end{thebibliography}
\end{document}